\documentclass[aps,pra,reprint,superscriptaddress,twocolumn]{revtex4-1}

\usepackage{graphicx}
\usepackage{dcolumn}
\usepackage{bm}
\usepackage{multirow}
\usepackage{upgreek}
\usepackage{bm}        
\usepackage{amssymb}   
\usepackage{amsmath}
\usepackage{epstopdf}
\usepackage{float}
\usepackage{color}
\usepackage[colorlinks,linkcolor=blue,citecolor=blue,urlcolor=blue]{hyperref}

\usepackage{subfig} 
\usepackage{caption}
\begin{document}

\title{Implementing Majorana fermions in a cold-atom honeycomb lattice with textured pairings}
\author{Ruizhi Pan}
\affiliation{Joint Quantum Institute and the University of Maryland, College Park, Maryland 20742, USA}
\author{Charles W. Clark}
\email{charles.clark@nist.gov}
\affiliation{Joint Quantum Institute and the University of Maryland, College Park, Maryland 20742, USA}
\affiliation{National Institute of Standards and Technology, Gaithersburg, Maryland 20899, USA}

\date{\today}

\begin{abstract}
Recent studies in the realization of Majorana fermion (MF) quasiparticles have focused on engineering topological superconductivity that derives from proximity effects of conventional superconductors and spin textures. We propose an effective model to create unpaired MFs at a honeycomb lattice edge by generalizing a 2-dimensional topologically nontrivial Haldane model and introducing textured pairings. The core idea is to add both the spin-singlet and textured spin-triplet pairings to a pseudospin-state dependent, time-reversal symmetry (TRS) noninvariant honeycomb lattice, and to satisfy generalized "sweet spot" conditions as in the Kitaev chain model. Our model has a gapped superconducting phase and a gapless phase; either phase may have zero or nonzero topological winding numbers. The discriminant that distinguishes those two phases gives a measure of TRS breaking and may have more general implications. Effective Majorana zero modes arise at edges in distinct phases with different degrees of degeneracy. Our theoretical model motivates concepts, such as "textured pairings" and the "strength" of TRS breaking, that may play important roles in future implementation of MFs with cold atoms in optical lattices.
\end{abstract}


\maketitle

\section{Introduction}

Majorana fermions (MF) have attracted much attention in recent years due to their implications for particle physics and potential applications to fault-tolerant topological quantum computation~\cite{MFnuclear,alicea2011non,sarma2015majorana}. Many protocols for the realization of MFs and implementing nonabelian statistics have been proposed, yet no single platform has been identified to be ideal for studies in all aspects~\cite{MFphasetrans,MFEquil,1DTOPChain,MFquasi1D,MFvort,zerobias,propobserv,MFferr,obMFferr}. In versatile platforms, such as condensed matter and quantum gas systems, MFs arise as Bogoliubov quasiparticle excitations at the defect sites (vortices, interfaces, system edges, etc.).
Some recently studied systems are related to topological superconductors that derive from proximity effects of conventional superconductors and spin textures~\cite{MFferr,MFpwaveSF,SCprox,obMFferr,highres,contrfinite,localoscil,majquasi}. Theoretical models of MFs in electronic materials originate from
p-wave pairing states of fermions with broken parity and time-reversal symmetry (TRS)~\cite{pairstate}. Much attention is given to techniques for detection and control of MFs, such as the preparation of spin-triplet pairing in p-wave superconductors, and to obviation of the need for precise parameter tuning~\cite{obMFferr,highres,mixtrip,tunepair,PrepProb}.

Implementations of MFs with cold atoms in optical lattices is also of interest~\cite{PrepProb,1DTOPChain,MFquasi1D,MFpwaveSF,TOPquanmatter,Createsingmaj}. This possibility is based on progress in creating topological phases of cold atoms using the development of synthetic spin-orbit coupling (SOC) and magnetic fields, s-wave and p-wave superfluidity and single-site addressing techniques, etc~\cite{1DTOPChain,MFEquil,MFquasi1D,zhang2008p,MeasureTOP,exprealiz,TImetal,designlaser,liu2016detecting}. In optical lattices, most theoretical models begin with Kitaev's 1-dimensional (1-D) p-wave superconducting (SC) quantum wire model~\cite{unpairMF} with SOCs, and extend it to 2-D using multiple parallel chains with interchain couplings~\cite{MFEquil,1DTOPChain,MFquasi1D,fateheir}. Such approaches yield
single or multiple 1-D topologically non-trivial chains in a background of trivial higher-dimensional optical lattices; isolated MFs emerge at chain ends in an odd-number-chain phase. This exotic topology still originates from the 1-D Kitaev model, while weak transverse tunneling suppresses quantum fluctuations and stabilizes the long-range order~\cite{MFquasi1D}. It is desirable to identify schemes that naturally include tunneling in different directions and incorporate the techniques in topological fermionic optical lattices to advance research on MFs.

In this paper, we propose an effective model to create MFs at an edge of the honeycomb lattice by introducing textured pairings into a 2-D topologically nontrivial Haldane model~\cite{MeasureTOP}. The key idea is to incorporate both the spin-singlet and textured spin-triplet pairings in the pseudospin-state dependent honeycomb optical lattice which breaks the TRS with complex next-nearest-neighbor (NNN) hopping. By tuning the pair coupling strength to match the amplitude and phase of hopping terms, MFs with flat bands (also called Majorana zero mode, "MZM") will arise on a single edge of the lattice.
This is similar to the "sweet spot" conditions in Kitaev model. This suggests that to realize such MZMs, it is critical to break the 3-fold rotational symmetry of the  pairing terms of the Hamiltonian, leading to a specific type of Majorana coupling. This requirement on the angular dependence of the sign of the spin-triplet pairing term is reminiscent of textured pairings in paired states of fermions~\cite{pairstate}.

In our model, the cold atom system has a gapped SC phase and a gapless phase for parameters in the "sweet spot". Either phase can have a winding number $w=0$ or $w\neq 0$. The phase diagram can be represented in the domain of phase parameters in the complex NNN hoppings. We demonstrate a method to reduce the gap-closing condition of the bulk Hamiltonian to the calculation of the discriminant.
This circumvents the analytical complexity of a 4-band model. The value of this discriminant distinguishes the gapped SC and gapless phases. It actually measures the  "strength" of TRS breaking, thus further dividing the TRS-broken class into two groups. In the gapped SC phase, there always exist two pairs of MZMs while the winding number of bulk bands, $w=\pm 1$, is associated with extra normal gapless edge states. One of the MZM pairs can be fully pseudospin-polarized localized at an edge in special cases, while the other pair usually extends to deeper layers with exponentially decaying amplitudes. In the gapless phase, the second pair of MZMs vanishes due to their coupling with the bulk modes. It remains to be determined whether the two pairs of MZMs in topological trivial cases will have an energy splitting in an extended model that incorporates the interaction of MFs or other coupling channels~\cite{stronginter}.

This paper is organized as follows: In section $\textrm {\uppercase\expandafter{\romannumeral 2}}$, we introduce our theoretical model and the intuition of generating MZMs. In section $\textrm {\uppercase\expandafter{\romannumeral 3}}$, we identify the MZMs from the aspects of the band structure, density profile and wavefunction symmetry by numerical simulation. In section $\textrm {\uppercase\expandafter{\romannumeral 4}}$, the phase diagram of a cold atom system is presented and the degeneracy of MZMs in each phase is discussed. We describe a mathematical method to find a discriminant that gives the phase boundary between gapped SC and gapless phases. This discriminant characterizes the "strength" of TRS breaking. Finally, our model is compared with previous models of creating MFs in 2-D cold atom systems.

\section{Model and physical intuition}

\captionsetup[subfigure]{position=top,singlelinecheck=off,justification=raggedright}
\begin{figure}[tbp]
\centering
 \subfloat[] {\includegraphics[width=0.5\textwidth]{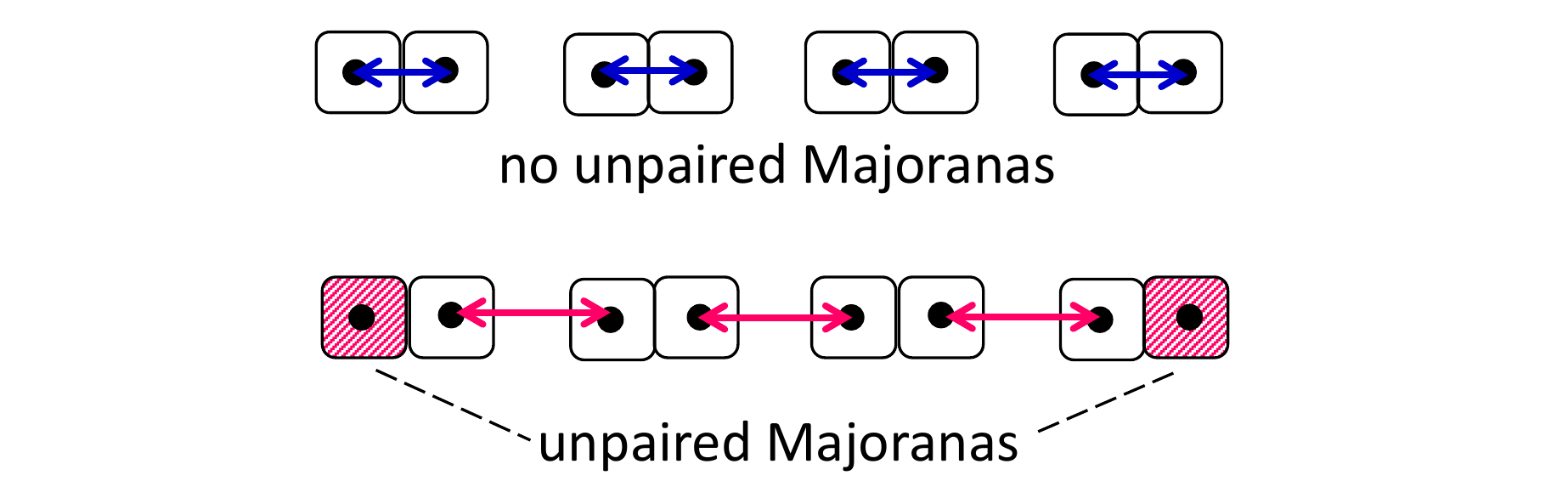} \label{fig1a}}\\

 \subfloat[] {\includegraphics[width=0.5\textwidth]{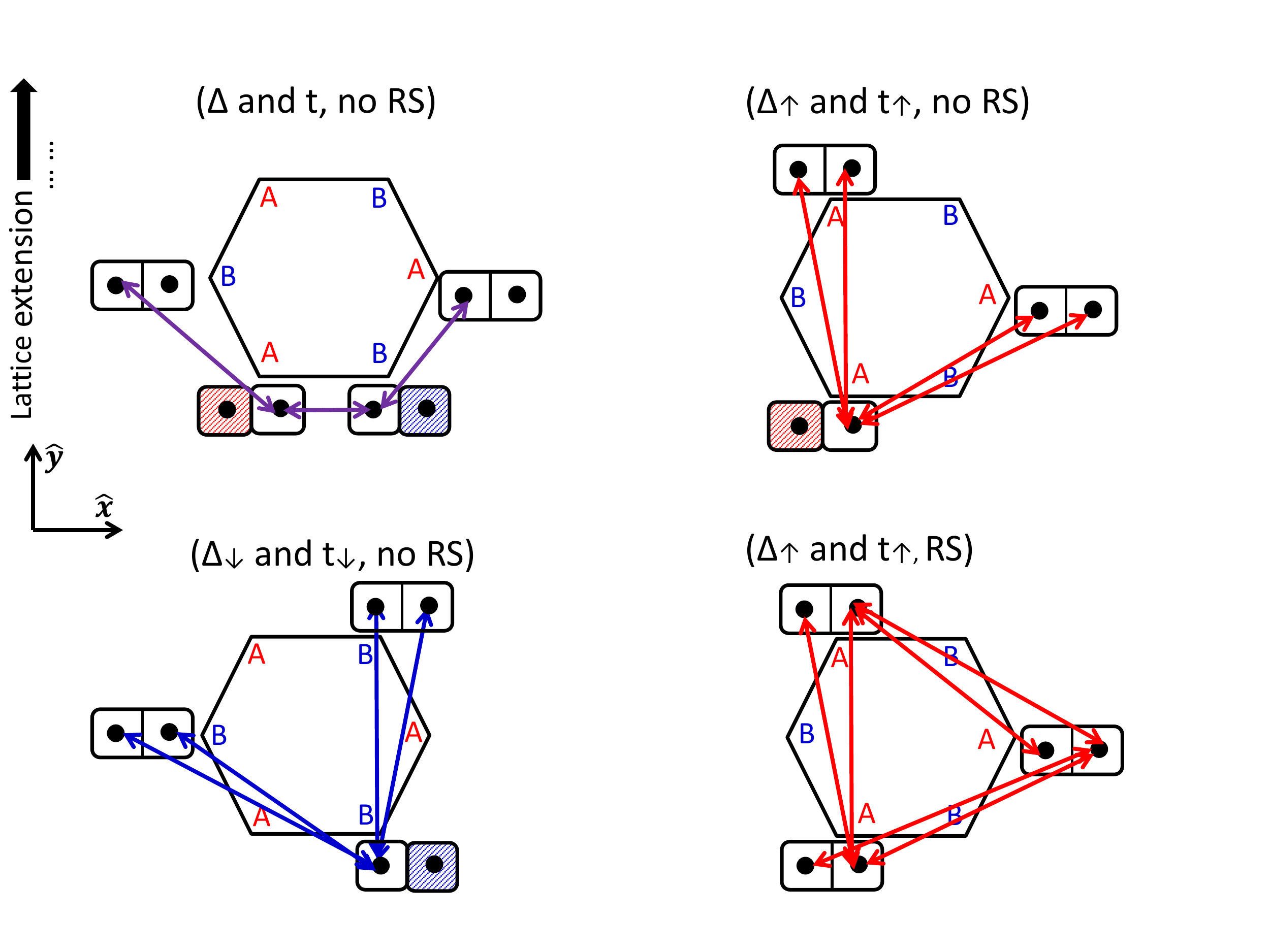} \label{fig1b}}\\
 \captionsetup{justification=raggedright}
 \caption{Physical intuition of generating unpaired MFs at an edge in the case of $\mu=0$. (a) Kitaev's 1-D spinless p-wave SC quantum wire. The two neighboring MFs constitute a normal fermion. The blue arrows in the upper chain signify the internal pairing of MFs with no unpaired MFs remaining. The red arrows in the lower chain indicate the inter-cell pairing of MFs with two unpaired MFs at the ends of the chain. (b) MF coupling at a single armchair edge of a 2-D honeycomb lattice. The upper two and lower left subfigures are for our model in which there is no 3-fold rotational symmetry (RS). The solid bonds are the net Majorana couplings contributed by terms related to $(\Delta, t)$, $(\Delta_{\uparrow}, t_{\uparrow})$ and $(\Delta_{\downarrow}, t_{\downarrow})$ in $\hat{H}$, respectively. The shaded and colored cells denote the dangling MFs in a hexagon at a single armchair edge of the lattice. The lower right subfigure shows an example of unexpected MF couplings in which there is rotational symmetry, just in comparison with that in our model.}
\end{figure}

Our model is based on a generalized Haldane model in a pseudospin-state dependent honeycomb optical lattice~\cite{MeasureTOP}, which is among many protocols proposed to implement the topological phases in systems of noninteracting fermions. In the realization of our system, ultracold atoms with two different hyperfine states would be described as two pseudospin states (spin-up "$\uparrow$" and spin-down "$\downarrow$"), each localized at one of two inequivalent sublattices (A and B). The natural tunneling between sites on the same sublattice and the laser-induced coupling between different sublattices implement the next-nearest-neighbor (NNN) and nearest-neighbor (NN) hoppings, respectively. To generate unpaired MFs, we add spin-dependent pair interactions between atoms to obtain the total effective Hamiltonian:

\begin{equation}
\hat{H}=\hat{H}_{0}+\hat{H}_{p},
\label{eq1}
\end{equation}
where
\begin{eqnarray}
\hat{H}_{0}=-t\sum_{\langle j,m\rangle }(a_{\vec{r}_{j}}^{\dagger} b_{\vec{r}_{m}}+h.c.)-t_{\uparrow}\sum_{\langle \langle j,j^{\prime}\rangle \rangle }(e^{i\phi_{A}} a_{\vec{r}_{j}}^{\dagger} a_{\vec{r}_{j^{\prime}}}+h.c.)  \nonumber \\
-t_{\downarrow}\sum_{\langle \langle m,m^{\prime}\rangle \rangle }(e^{i\phi_{B}} b_{\vec{r}_{m}}^{\dagger} b_{\vec{r}_{m^{\prime}}}+h.c.) \nonumber \\
+\mu(\sum_{j}a_{\vec{r}_{j}}^{\dagger} a_{\vec{r}_{j}}-\sum_{m}b_{\vec{r}_{m}}^{\dagger} b_{\vec{r}_{m}}),  \nonumber  \\
\\
\label{eq2}
\hat{H}_{p}=\sum_{\langle j,m\rangle }(\Delta a_{\vec{r}_{j}}^{\dagger} b_{\vec{r}_{m}}^{\dagger}+h.c.)
+\sum_{\langle \langle j,j^{\prime}\rangle \rangle ,y_{j}< y_{j^{\prime}}}(\Delta_{\uparrow} a_{\vec{r}_{j}}^{\dagger}a_{\vec{r}_{j^{\prime}}}^{\dagger}+h.c.) \nonumber \\
+\sum_{\langle \langle m,m^{\prime}\rangle \rangle ,y_{m}< y_{m^{\prime}}}(\Delta_{\downarrow} b_{\vec{r}_{m}}^{\dagger}b_{\vec{r}_{m^{\prime}}}^{\dagger}+h.c.). \nonumber \\   
\label{eq3}
\end{eqnarray}

In the equations above, $\hat{H}_{0}$ is the original effective Hamiltonian following a unitary basis transformation as described in Ref.~\cite{MeasureTOP}; $t$ is the NN hopping amplitude and $t_{\uparrow}$($t_{\downarrow}$) is the NNN hopping amplitude in sublattice A(B). These three hopping amplitudes are real numbers. The NNN hopping phases
are given by $\phi_{A}$($\phi_{B}$), and $\vec{r}_{j}$($\vec{r}_{m}$) denotes the site index of sublattice A(B) (note that there's only one spin state at one site so that $\vec{r}_{j}$ and $\vec{r}_{m}$ are from different displacement vector sets). $\mu$ denotes half of the chemical potential difference between sublattice A and B. $\hat{H}_{p}$ is the spin-dependent pairing Hamiltonian introduced by our model. $\Delta$ terms denote the pairings of atoms with opposite spins (spin-singlet), while $\Delta_{\uparrow}$ and $\Delta_{\downarrow}$ terms denote the pairings of atoms with the same spins (spin-triplet).

For simplicity, we first analyze a special case in which the on-site staggered potential vanishes, i.e., $\mu$ is zero. Then, the system can be viewed in the Majorana representation (Fig. \ref{fig1b}) by rewriting the Hamiltonian with Majorana operators:
\begin{eqnarray}
a_{\vec{r}_{j}}=\frac{1}{2}(\gamma_{\vec{r}_{j},\uparrow,1}+i\gamma_{\vec{r}_{j},\uparrow,2}),
b_{\vec{r}_{m}}=\frac{1}{2}(\gamma_{\vec{r}_{m},\downarrow,1}+i\gamma_{\vec{r}_{m},\downarrow,2}). \nonumber
\end{eqnarray}
Here, $\{\gamma_{\vec{r}_{j},\sigma,\alpha},\gamma_{\vec{r}_{m},\sigma^{\prime},\beta}\}=2\delta_{\vec{r}_{j},\vec{r}_{m}}
\delta_{\sigma\sigma^{\prime}}\delta_{\alpha\beta}$ with each Majorana operator, $\gamma$, having three subscripts denoting the position, spin and Majorana type, respectively. In the following discussions, a Majorana coupling usually means a $i\gamma\gamma^{\prime}$ term in the Hamiltonian and is denoted by a bond (double arrow) between two sites in Fig. \ref{fig1a} and \ref{fig1b}. The Majorana coupling
contributes an ordinary fermion which costs the energy of a bulk mode in the band structure. Based on the physical intuition shown in Fig. \ref{fig1b}, we choose to introduce $\hat{H}_{p}$ in the above form, in order to cancel a part of the Majorana couplings introduced by the $t$, $t_{\uparrow}$ and $t_{\downarrow}$ hopping interactions. This choice yields the Hamiltonian in Appendix A.

The new "sweet spot" conditions that create dangling MFs at edges can be deduced. We inherit the key idea of Kitaev's 1-D spinless p-wave SC chain model, which is to choose a specific type of Majorana coupling and leave unpaired MFs at the edges of the finite lattice (Fig. \ref{fig1a}). In our model, we choose to make the net effect of $\Delta$ and $t$ terms be the coupling $\gamma_{\vec{r}_{j},\uparrow,2}\gamma_{\vec{r}_{m},\downarrow,1}$ between NN sites, and the net effect of $\Delta_{\uparrow(\downarrow)}$ and $t_{\uparrow(\downarrow)}$ terms be the two couplings $\gamma_{\vec{r}_{j},\uparrow(\downarrow),2(1)}\gamma_{\vec{r}_{j^{\prime}},\uparrow(\downarrow),\alpha}$ between NNN sites with $y_{j}<y_{j^{\prime}}$ and $\alpha=1,2$ (Fig. \ref{fig1b}), where $y_{j}$ and $y_{j^{\prime}}$ are the $\hat{y}$ component of $\vec{r}_{j}$ and $\vec{r}_{j^{\prime}}$, respectively. It means that the coupling connecting A and B sublattices is only between type-2 MFs in A and type-1 MFs in B. And the coupling within A(B) sublattice is only from type-2(1) MFs to MFs of both two types with bigger indices in $\hat{y}$ direction. These requirements can be reduced to Eq. (\ref{eq4})-(\ref{eq6}) which can be called the "sweet spot" in the parameter space. So the final result is that the type-1 MFs of atoms in A sublattice and type-2 MFs of atoms in B sublattice at one armchair edge (the shaded and colored cells in Fig. \ref{fig1b}) are isolated, i.e., $\gamma_{y_{1},\uparrow,1}$ and $\gamma_{y_{1},\downarrow,2}$ don't appear in $\hat{H}$, as there's no other MFs outside the lattice to couple with them.\\

\begin{eqnarray}
\Delta&=&-t,
\label{eq4}\\
\Delta_{\uparrow}&=&-t_{\uparrow}e^{-i\phi_{A}},
\label{eq5}\\
\Delta_{\downarrow}&=&t_{\downarrow}e^{-i\phi_{B}}.
\label{eq6}
\end{eqnarray}

\captionsetup[subfigure]{position=top,singlelinecheck=off,justification=raggedright}
\begin{figure}[tbp]
\centering
\subfloat[] {\includegraphics[width=0.20\textwidth]{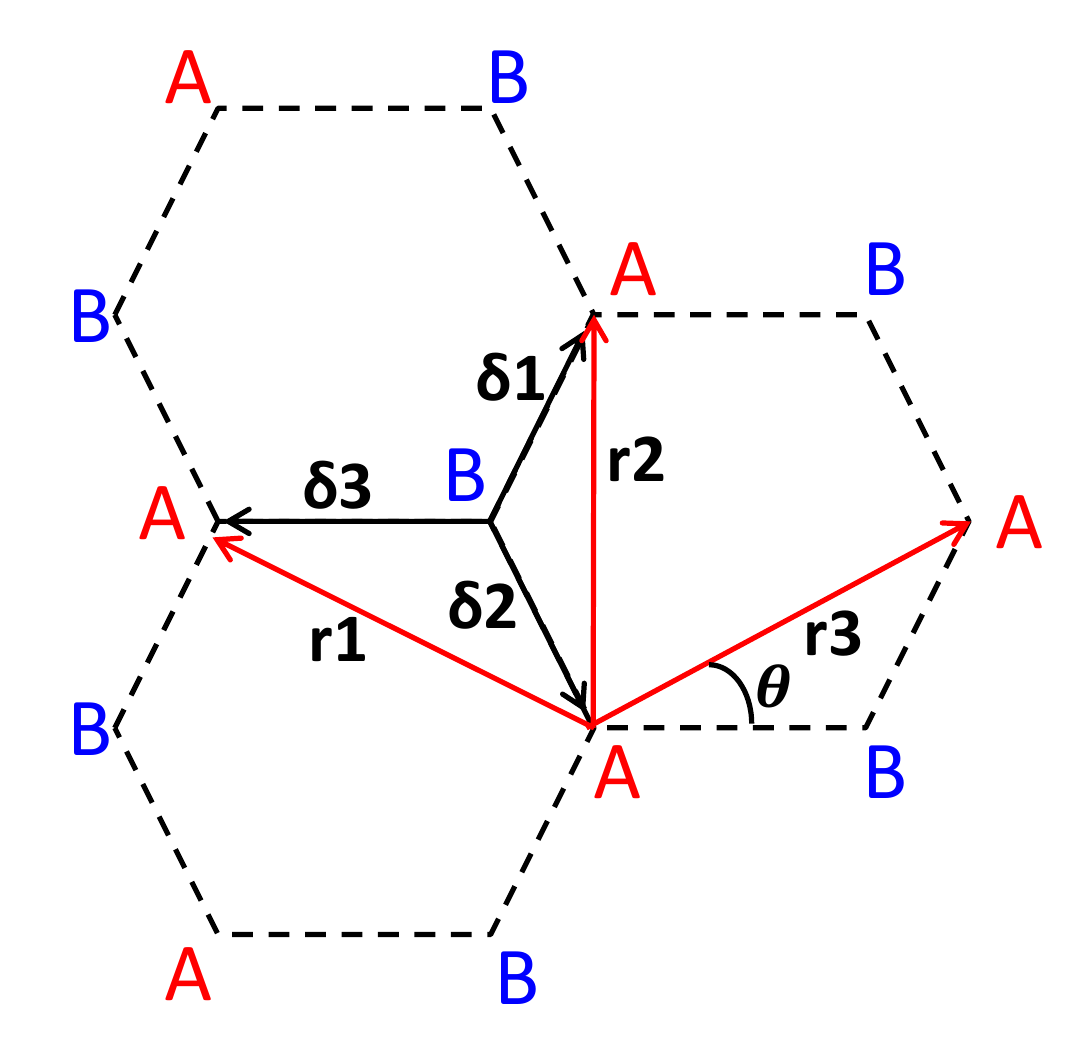} \label{fig2a}}
\subfloat[] {\includegraphics[width=0.25\textwidth]{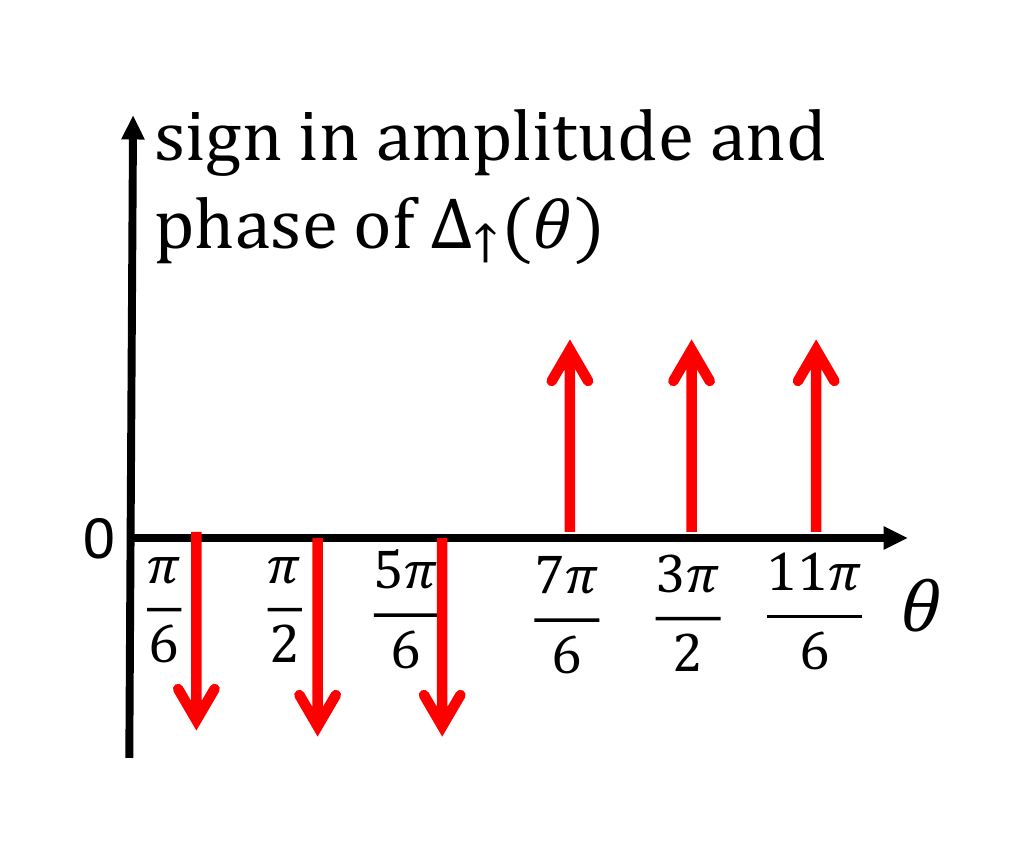} \label{fig2b}}\\

\subfloat[] {\includegraphics[width=0.5\textwidth]{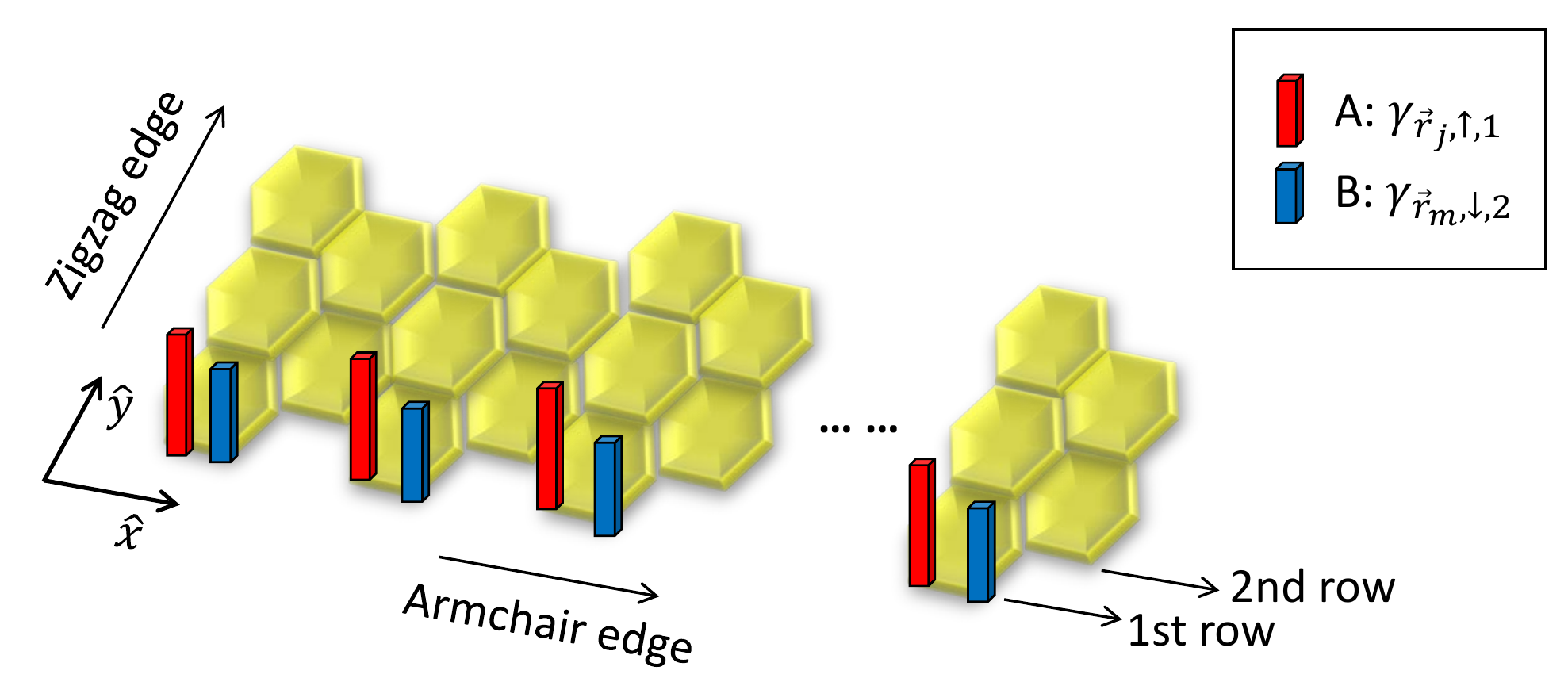} \label{fig2c}}
\captionsetup{justification=raggedright}
\caption{Schematic of our physical system. (a) The depictions of the three vectors ($\vec{\delta}_{j}$) connecting NN sites and the three vectors ($\vec{r}_{j}$) connecting NNN sites for $j=1,2,3$. (b) Angular distribution of the sign in the amplitude and phase of the defined pairing which is similar to the domain wall structure. (c) The distribution of MZMs at a single armchair edge. The heights of the pillars denote the amplitudes of wavefunction at each site.}
\end{figure}

Therefore, our model has generalized "sweet spot" conditions analogous to those of a 1-D Kitaev chain, and actually possesses textured pairings analogous to that in the original model of fermionic paired states~\cite{pairstate}. Note first that the $\Delta_{\uparrow(\downarrow)}$ terms are written in a particular order ($y_{j(m)}< y_{j^{\prime}(m^{\prime})}$) to avoid mixing definitions. Second, it's critical to break the 3-fold rotational symmetry of the net Majorana couplings within a hexagon (Fig. \ref{fig1b}). Since $t$ is real, the net coupling between NN A and B sites may be reduced to one bond (the upper left subfigure in Fig. \ref{fig1b}). However, there's no degree of freedom to reduce the net coupling between NNN A or B sites to less than 2 bonds since $t_{\sigma}$ is complex. Thus, any pairing Hamiltonian that has 3-fold rotational symmetry (such as the lower right subfigure in Fig. \ref{fig1b}) doesn't allow dangling MFs at edges. We need to impose requirements on the sign in the amplitude and phase of $\Delta_{\sigma}$ to break the 3-fold rotational symmetry.

We present the requirement on pairing terms in an alternative way, which highlights the concept of textures in pairing. Following the reference~\cite{MeasureTOP}, choose the Peierls phase associated with the NNN hopping path $a_{\vec{r}_{j}}^{\dagger} a_{\vec{r}_{j^{\prime}}}$ to be:
\begin{equation}
\phi_{A}(j,j^{\prime})=-\vec{p}\cdot (\vec{r}_{j}-\vec{r}_{j^{\prime}})/2. \nonumber
\end{equation}
A pair creation term in $\hat{H}$ is $\Delta_{\uparrow}(\theta)a_{\vec{r}_{j}}^{\dagger} a_{\vec{r}_{j^{\prime}}}^{\dagger}$, where $\theta$ is the angle between $(\vec{r}_{j^{\prime}}-\vec{r}_{j})$ and the $\hat{x}$ axis. Then, we can deduce from the above requirements:

\begin{displaymath}
\Delta_{\uparrow}(\theta)= 
\left\{
\begin{array}{l} 
-t_{\uparrow}e^{-i\vec{p}\cdot (\vec{r}_{j^{\prime}}-\vec{r}_{j})/2}, \hspace{0.3in} (y_{j}<y_{j^{\prime}})  \\
t_{\uparrow}e^{i\vec{p}\cdot (\vec{r}_{j^{\prime}}-\vec{r}_{j})/2}. \hspace{0.52in} (y_{j}>y_{j^{\prime}}) \\
\end{array}
\right.  
\end{displaymath}

We can see that the sign in $\Delta_{\uparrow}(\theta)$ in front of the amplitude $t_{\uparrow}$ and phase $\vec{p}\cdot (\vec{r}_{j^{\prime}}-\vec{r}_{j})/2$ is changed across $[0, 2\pi]$ (Fig. \ref{fig2b})). This requirement of the broken rotational symmetry leads to exotic textures in the pairing terms~\cite{pairstate}, and the angular distribution of the sign in the spin-triplet pairing term has a reorientation similar to that of the domain wall structure in magnetism~\cite{catalan2012domain}. Following the concept of spin textures strongly related to the topological superconductivity, such as the vortex, skyrmion, spiral and helix, this discrete texture in the pairing we just described is a generalization and may play an important role in future studies of MFs in honeycomb lattice structures.

It should be noted that we get our intuition from the particular case where "$\mu=0$", but the effectiveness of our model in creating MFs is not limited to this case. The case where "$\mu \neq 0$" is discussed in section $\textrm {\uppercase\expandafter{\romannumeral 4}}$ below. Also note that we just give one particular type of MF couplings. It determines the location of unpaired MFs which are not necessarily at the armchair edge of the lattice.

\section{Identification of Majorana zero modes}

\captionsetup[subfigure]{position=top,singlelinecheck=off,justification=raggedright}
\begin{figure}[tbp]
\centering
 \subfloat[Gapped SC, $w=0$] {\includegraphics[width=0.25\textwidth]{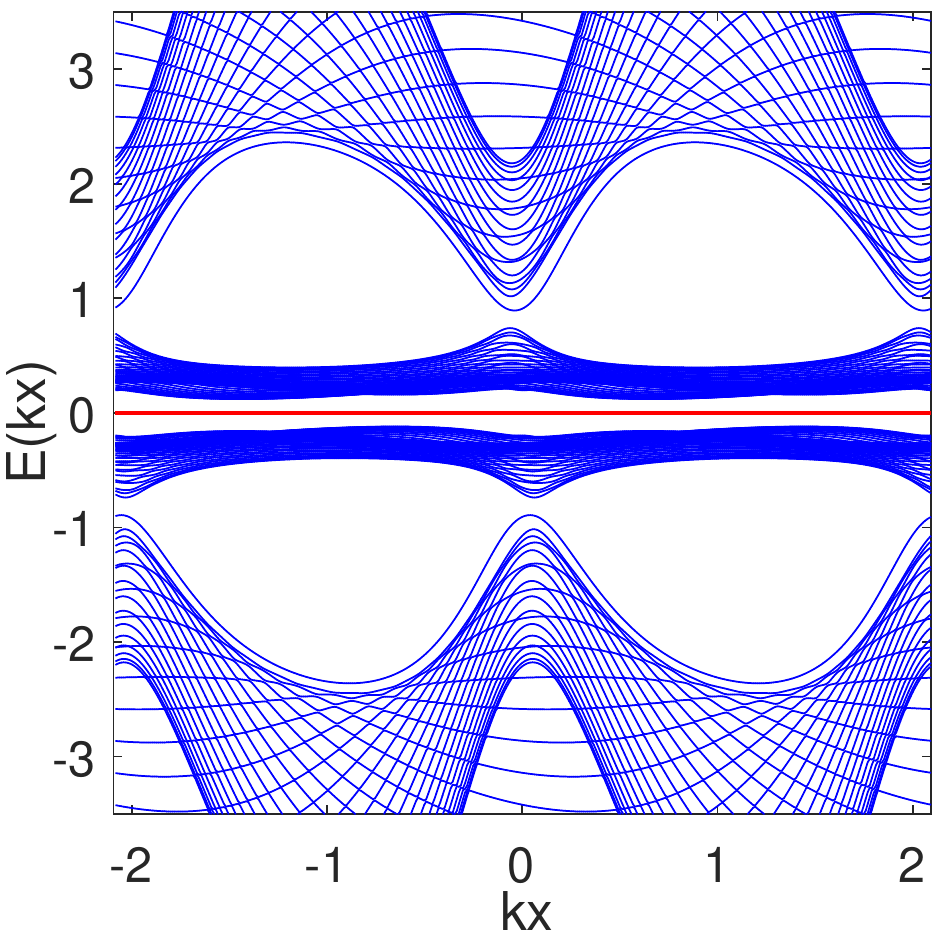} \label{fig3a}}
 \subfloat[Gapped SC, $w=\pm 1$] {\includegraphics[width=0.25\textwidth]{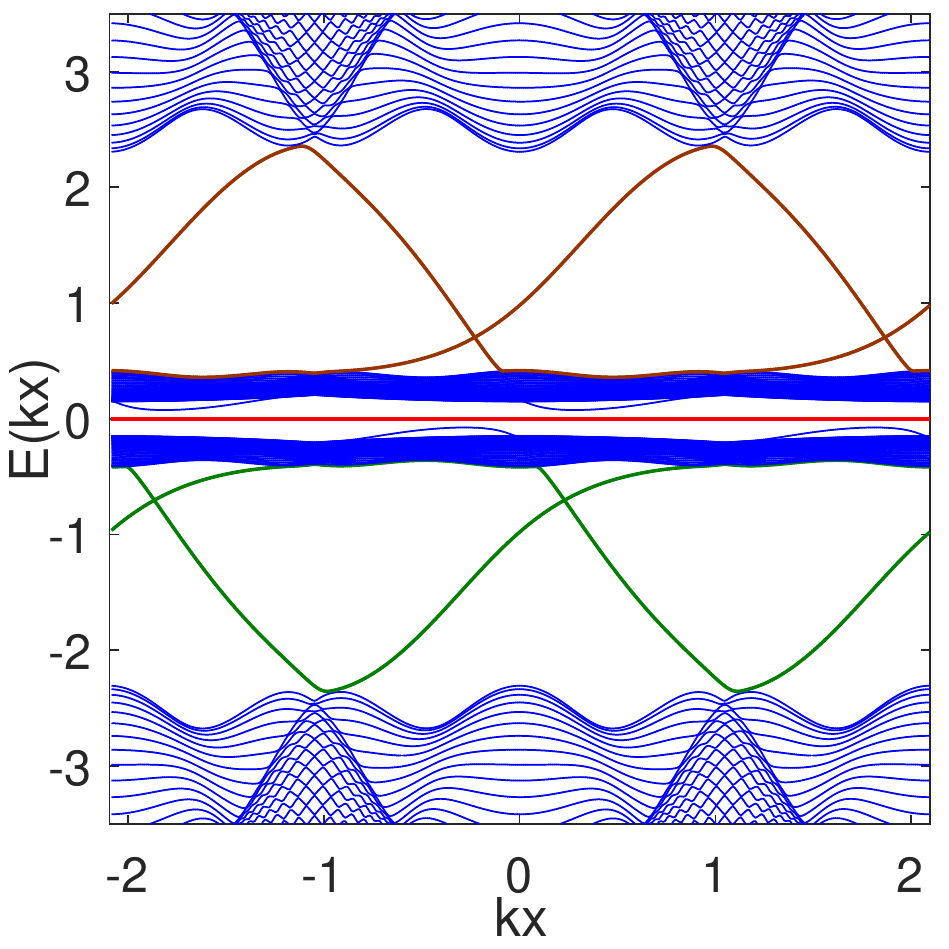} \label{fig3b}}\\

 \subfloat[Gapless, $w=0$] {\includegraphics[width=0.25\textwidth]{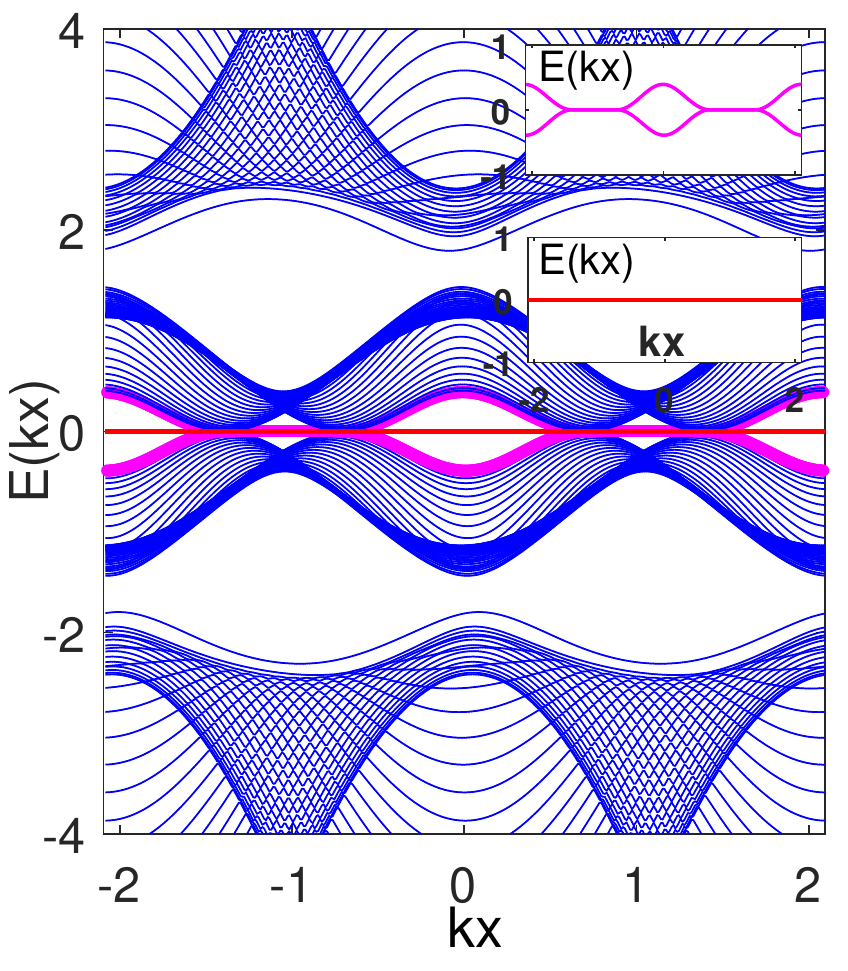} \label{fig3c}}
 \subfloat[Gapless, $w=\pm 1$] {\includegraphics[width=0.25\textwidth]{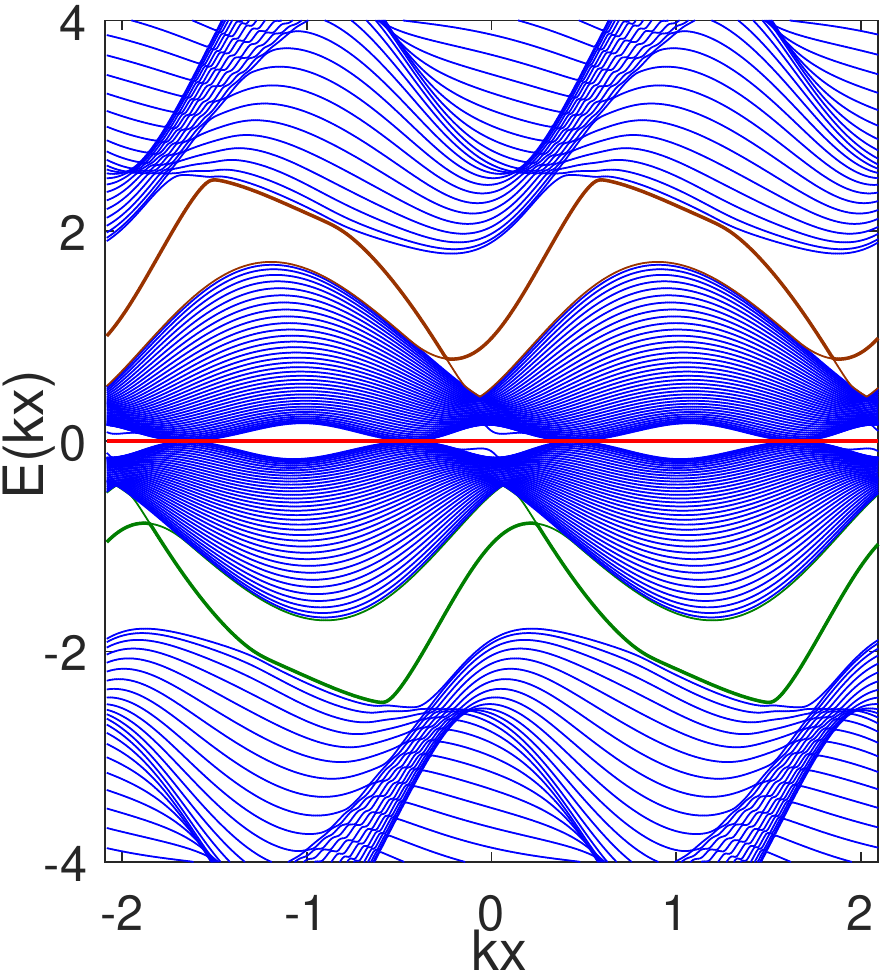} \label{fig3d}}
 \captionsetup{justification=raggedright}
 \caption{Band structure for $\mu=0$ in four cases characterized by different gap conditions and winding numbers $w$. Fixed parameters: $t=1$, $t_{\uparrow}=0.4$, $t_{\downarrow}=0.6$. (a) $\vec{p}=(0.9K_{x},0)$; (b) $\vec{p}=(0,3K_{y})$; (c) $\vec{p}=(0.1K_{x},0.2K_{y})$; (d) $\vec{p}=(0.6K_{x},3K_{y})$, where $K_{x}=2\pi/3$ and $K_{y}=K_{x}/\sqrt{3}$. Each group of blue curves represents a bulk band. The dark red straight line is the MZM and the brown and green curves in (b) and (d) are gapless edge modes. In (a) and (b), the zero modes are 4-fold degenerate; in (c) and (d), the zero modes are 2-fold degenerate. The inset in (c) zooms in on the split zero modes (magenta) and flat MZMs (dark red) separately.}
\end{figure}

We can identify MZMs from the aspects of the band structure, density profile and wavefunction symmetry obtained by numerical simulations. The geometric definition of our model is depicted in Fig. \ref{fig2a}. We use three displacement vectors $\vec{\delta}$ to denote the NN hoppings and another three $\vec{r}$ to denote the NNN hoppings:

\begin{eqnarray}
\vec{\delta}_{1}=a(\frac{1}{2},\frac{\sqrt{3}}{2}),
\vec{\delta}_{2}=a(\frac{1}{2},-\frac{\sqrt{3}}{2}),
\vec{\delta}_{3}=a(-1,0), \nonumber \\
\vec{r}_{1}=a(-\frac{3}{2},\frac{\sqrt{3}}{2}),
\vec{r}_{2}=a(0,\sqrt{3}),
\vec{r}_{3}=a(\frac{3}{2},\frac{\sqrt{3}}{2}). \nonumber
\end{eqnarray}

\noindent Here, $a$ is the side length of a hexagon plaquette and is set as the unit of length in this paper. Using conventions of current techniques of implementing the complex NNN hopping by laser-induced transitions~\cite{MeasureTOP}, we choose the Peierls phase associated with $a_{\vec{r}_{j}}^{\dagger} a_{\vec{r}_{j^{\prime}}}$ to be $\phi_{A}(j,j^{\prime})=-\vec{p}\cdot (\vec{r}_{j}-\vec{r}_{j^{\prime}})/2$ and similarly $\phi_{B}(m,m^{\prime})=\vec{p}\cdot (\vec{r}_{m}-\vec{r}_{m^{\prime}})/2$, where $\vec{p}$ is the momentum transfer associated with the laser-induced tunneling.

We calculate the band structure using a momentum space representation based on a Fourier transformation in the $\hat{x}$ direction:
\begin{eqnarray}
\hat{a}_{\vec{r}_{j}}&=&\frac{1}{\sqrt{N_{x}}}\sum_{k_{x}}e^{i k_{x}x_{j}}\hat{a}_{k_{x},y_{j}},  \nonumber \\ \hat{b}_{\vec{r}_{m}}&=&\frac{1}{\sqrt{N_{x}}}\sum_{k_{x}}e^{i k_{x}x_{m}}\hat{b}_{k_{x},y_{m}}.  \nonumber
\end{eqnarray}
$x_{j(m)}$ and $y_{j(m)}$ are the components of $\vec{r}_{j(m)}$ in $x$ and $y$ direction, respectively. $N_{x}$ is the number of cells along $\hat{x}$ direction, much larger than that along $\hat{y}$ direction. Thus the basis vector of the Bogoliubov-de-Gennes Hamiltonian of $\hat{H}$ for particular $k_{x}$ becomes $(\hat{a}_{k_{x},y_{1}}, ..., \hat{b}_{k_{x},y_{1}}, ..., \hat{a}^{\dagger}_{-k_{x},y_{1}}, ..., \hat{b}^{\dagger}_{-k_{x},y_{1}}, ...)^{T}$, in which the subscripts denoting $y$ component range over all the rows (also called layers in this paper). Then, the band structure containing both the edge modes and bulk bands is obtained by the diagonalization of the Hamiltonian in this basis. The results are shown in figures \ref{fig3a}-\ref{fig3d}.

Our model has a gapless and a gapped SC phase at the "sweet spot", either of which may have zero or nonzero winding numbers of the first excited band. As a common feature, there are two groups of bulk bands (blue curves) in each of the upper or lower half-planes. These are inherited from the Haldane model due to the number of inequivalent sites in a unit cell. The gapped SC phase and gapless phase are distinguished by the gap closing condition between the zero-energy line and the first excited band in the upper half-plane ("band 1"). In the gapped SC phase (Fig. \ref{fig3a}, \ref{fig3b}), there are two pairs of MZMs with complete flat bands. These are shown by straight red lines in Fig. \ref{fig3a}-\ref{fig3d}, coinciding with each other. By contrast, the gapless phase holds one pair of MZMs while the other pair of MZMs partially merges into the bulk modes in some ranges of $k_{x}$ (Fig. \ref{fig3c}, \ref{fig3d}). These "partial" MZMs that terminate at band closing points are lower dimensional Majorana analogues of Fermi arcs in 3D Weyl semi-metals~\cite{fateheir}. Furthermore, each of the gapped SC and gapless phases can have winding numbers  $w=0$ or $w=\pm 1$ of the band 1. The nonzero-winding-number phase corresponds to additional ordinary gapless edge modes between the band 1 and the second excited bulk band in the upper half-plane (band 2).

To further support the correctness of our model, we get the wavefunction of the zero modes in the band structure as below by numerical simulation:

\begin{eqnarray}
\hat{\Psi}_{k_{x}}=\sum_{y_{j}}(u_{k_{x},y_{j}}, v_{k_{x},y_{j}}, u_{k_{x},y_{j}}, -v_{k_{x},y_{j}})\cdot \nonumber \\
(\hat{a}_{k_{x},y_{j}},
\hat{b}_{k_{x},y_{j}}, \hat{a}^{\dagger}_{-k_{x},y_{j}}, \hat{b}^{\dagger}_{-k_{x},y_{j}})^{T} \nonumber \\
\approx \sum_{x_{j}}(\frac{1}{\sqrt{N_{x}}}u_{k_{x},y_{1}}e^{-i k_{x}x_{j}})\hat{\gamma}^{(A)}_{x_{j},y_{1},1} \nonumber \\
+\sum_{x_{m}}(\frac{1}{\sqrt{N_{x}}}v_{k_{x},y_{1}}e^{-i k_{x}x_{m}})\hat{\gamma}^{(B)}_{x_{m},y_{1},2},
\label{eq7}
\end{eqnarray}
where $u_{k_{x},y_{i}}$ and $v_{k_{x},y_{i}}$ are the wavefunctions for the $i$th layer in the $\hat{y}$ direction using the partial Fourier-transformed basis. Note that there is a degree of freedom in choosing the coefficient in front of a specific eigenstate, which ensures the above solution is a Majorana. We use the approximation sign and just keep the wavefunctions for the first layer in the above as the numerical simulation indicates that the solution amplitudes generally decay at an exponential rate (Fig. \ref{fig2c}).

The two pairs of MZMs in the gapped SC phase display some features. One pair can be fully pseudospin-polarized only localized at an edge layer when $\mu=0$. Such a pair has forms:
\begin{eqnarray}
(\hat{a}_{k_{x},y_{1}}+\hat{a}^{\dagger}_{-k_{x},y_{1}})|0\rangle, (\hat{b}_{k_{x},y_{1}}-\hat{b}^{\dagger}_{-k_{x},y_{1}})|0\rangle. \nonumber
\end{eqnarray}
It can be shown that they are always two zero-energy eigenvectors of the Bogoliubov-de-Gennes Hamiltonian of $\hat{H}$ in the partial Fourier-transformed basis, where $|0\rangle$ is the vacuum state. Both of these two MZMs persist in the gapless phase, similar to the persistence of edge MFs in a 1-D Kitaev chain with zero chemical potential (in which the Kitaev chain is always in a topologically nontrivial phase). The other pair of MZMs usually extends to deeper layers with exponentially decaying amplitudes. They have an energy splitting due to their coupling with bulk modes in the gapless phase. It should be noted that our model applies to noninteracting fermions and it suggests a new scheme for finding MFs. It remains to be determined by future research whether the two pairs of MZMs in topological trivial cases will have an energy splitting when the interaction between MFs or other coupling channels are added~\cite{stronginter}. If that is the case, there will be detectable effective MFs only in the $w\neq0$ phases.

\section{Phase diagram at the sweet spot}

\captionsetup[subfigure]{position=top,singlelinecheck=off,justification=raggedright}
\begin{figure}[tbp]
\centering
 \subfloat[Gapped SC vs. gapless]{\includegraphics[width=0.25\textwidth]{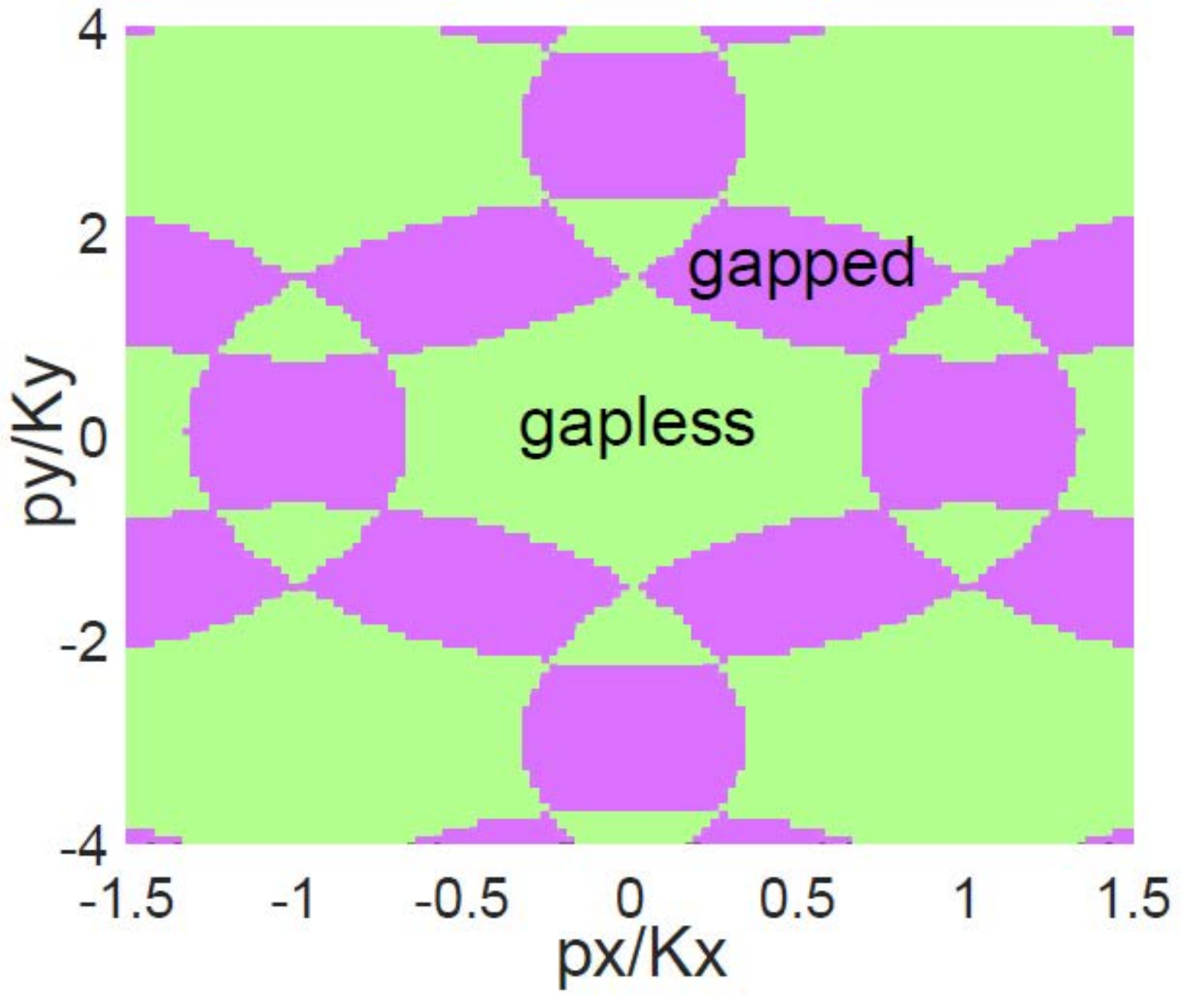} \label{fig4a}}
 \subfloat[$w=0$ vs. $w=\pm 1$] {\includegraphics[width=0.25\textwidth]{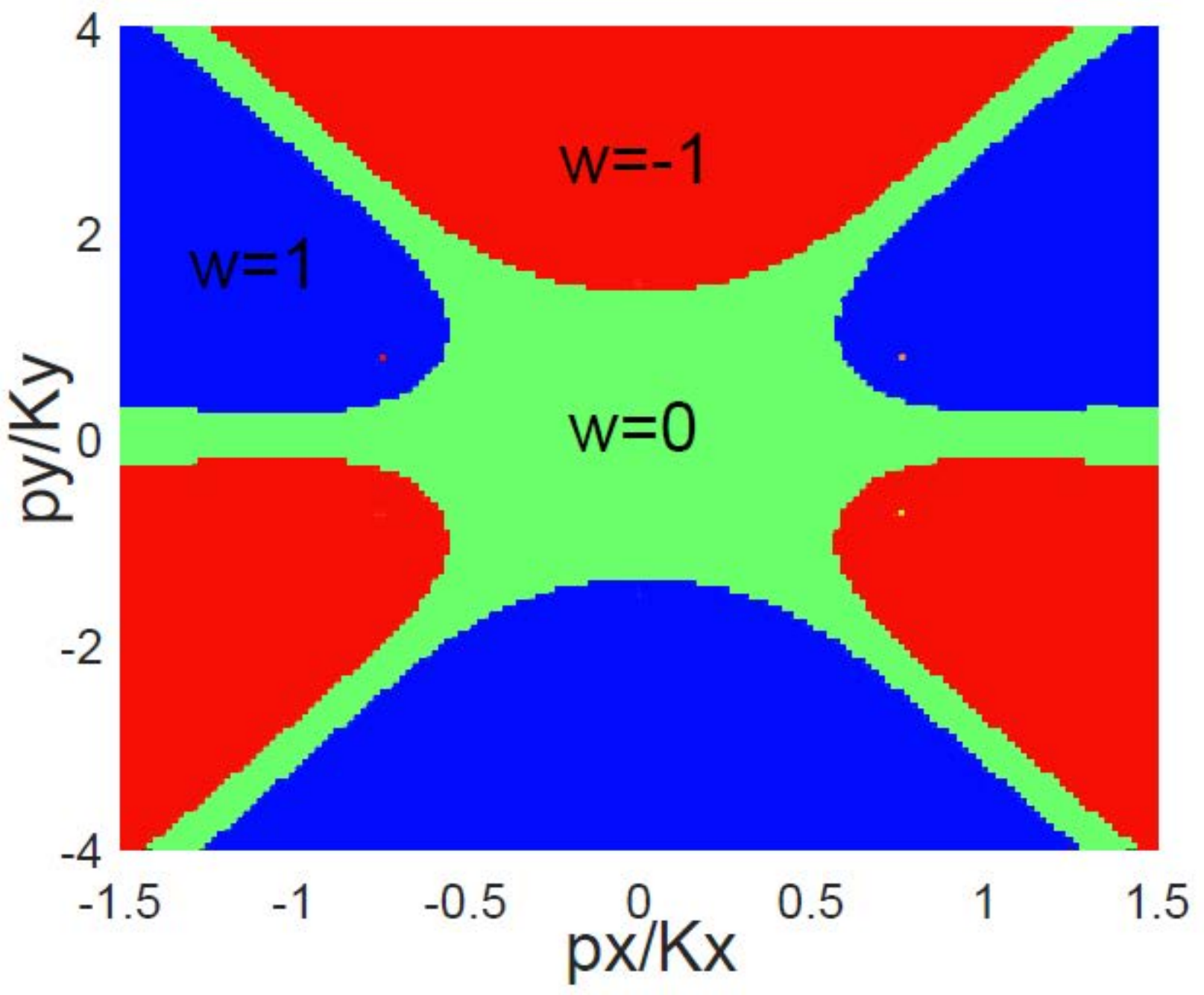} \label{fig4b}}\\
 \captionsetup{justification=raggedright}
 \caption{Phase diagram at the generalized "sweet spot" when $\mu=0$, obtained by numerical simulation. (a) shows the phase boundary between the gapped SC phase (purple) and gapless phase (light green). (b) shows the phase boundary between phases with different winding numbers. The green region indicates $w=0$, the blue region $w=1$ and the red region $w=-1$. A combination of these two figures shows the full 4-phase diagram.}
\end{figure}

The phase diagram of the cold atom system at the "sweet spot" is worth analyzing. For fixed amplitude parameters ($t$, $t_{\uparrow}$ and $t_{\downarrow}$), the phase of the system varies with the NNN hopping phases
$\phi_{A}$ and $\phi_{B}$. This is displayed by the phase diagram in the momentum coordinates $p_{x}$ and $p_{y}$, which is shown in Fig. \ref{fig4a}, \ref{fig4b}.
There are a total of four phases associated with the two alternatives "gapped SC vs. gapless" and "$w=0$ vs. $w=\pm 1$".

As we illustrated in section $\textrm {\uppercase\expandafter{\romannumeral 3}}$, the phase boundary between the gapped SC phase and gapless phase can be deduced by the bulk-edge correspondence in topological physics. In the above version of our model, the pseudospin space and the particle-hole space each contribute 2 degrees of freedom. Thus, we have a 4-band model, the description of which requires solutions of a quartic equation, which in general have complicated analytical forms. To circumvent the mathematical complexity of analytical band expressions of a 4-band model, we demonstrate a method to rigorously reduce into a discriminant the parameter conditions of the gap closing between band 1 and band 2 identified in section $\textrm {\uppercase\expandafter{\romannumeral 3}}$.

In the gapped SC phase, the gap between bands 1 and 2 is open due to TRS breaking by complex NNN hoppings, protecting the MZMs from coupling with bulk modes. In a certain range, this gap is approximately proportional to the amplitude of the relative complex hopping $|\frac{t_{\sigma}}{t}|$. At the "sweet spot", the Bogoliubov-de-Gennes Hamiltonian in momentum space, $H_{BdG}(\vec{k})$, reduces to:

\begin{displaymath}
\left( \begin{array}{cccc}
-2t_{\uparrow}f_{-}+\mu & -tg^{*} & |\Delta_{\uparrow}|h  & \Delta g^{*} \\
-tg & -2t_{\downarrow}f_{+}-\mu & -\Delta g & |\Delta_{\downarrow}|h^{*} \\
|\Delta_{\uparrow}|h^{*} & -\Delta g^{*} & 2t_{\uparrow}f_{+}-\mu & tg^{*} \\
\Delta g & |\Delta_{\downarrow}|h & tg & 2t_{\downarrow}f_{-}+\mu
\end{array} \right),
\end{displaymath}
where
\begin{eqnarray}
&&f_{+}=f_{+}(\vec{k},\vec{p})=\sum_{j=1}^{3}\cos((\vec{k}+\vec{p}/2)\cdot \vec{r_{j})}, \nonumber \\
&&f_{-}=f_{-}(\vec{k},\vec{p})=\sum_{j=1}^{3}\cos((\vec{k}-\vec{p}/2)\cdot \vec{r_{j})}, \nonumber \\
&&g=g(\vec{k})=\sum_{j=1}^{3}e^{-i\vec{k}\cdot \vec{\delta_{j}}}, \nonumber \\
&&h=h(\vec{k},\vec{p})=2i\sum_{j=1}^{3}e^{-i\frac{\vec{p}}{2}\cdot \vec{r_{j}}}\sin(\vec{k}\cdot \vec{r_{j}}). \nonumber
\end{eqnarray}

For a general quartic equation in $E$, $F(E)=\prod_{j=1}^{4}(E-E_{j})=0$, a two-fold root $E=0$, corresponding to band-touching conditions, implies the constraint $F(E=0)=\frac{dF}{dE}(E=0)=0$. The energy eigenvalues, $E$, can be calculated by solving
the characteristic polynomial $F(E)=\det(H_{BdG}(\vec{k})-E*I_{4\times4})=0$, where $\det()$ means the determinant. By invoking the "sweet spot" conditions Eq. (\ref{eq4})-(\ref{eq6}), $\mu=0$ and the band constraint, we get:

\begin{eqnarray}
F(E=0)&=&(t_{\uparrow}t_{\downarrow})^{2}(4f_{+}f_{-}+|h|^{2})^{2},
\label{eq8}
\\
\frac{dF}{dE}(E=0)&=&2t_{\uparrow}t_{\downarrow}(t_{\uparrow}-t_{\downarrow})(f_{+}-f_{-})(4f_{+}f_{-}+|h|^{2}).
\nonumber \\
\label{eq9}
\end{eqnarray}

We make three observations that lead to a fuller understanding of the solutions of the Eq. (\ref{eq8})-(\ref{eq9}). First, these equations contain a common nonnegative factor (discriminant)
\begin{equation}
4f_{+}f_{-}+|h|^{2}=4|\sum_{j=1}^{3}e^{i\vec{k}\cdot \vec{r}_{j}}\cos(\vec{p}\cdot \vec{r}_{j}/2)|^{2}.
\label{eq10}
\end{equation}
Thus, when  Eq. (\ref{eq8}) vanishes ($t_{\uparrow}t_{\downarrow}\neq 0$ typically), Eq. (\ref{eq9}) also vanishes. So there must be 0 or 2 bands (or more bands) simultaneously touching the zero-energy line. This is consistent with the fact that the two intermediate bulk bands touch with the zero-energy line at the same points in the band structure.

Second, the terms $t$ and $g(\vec{k})$ do not affect the band touching condition. Thus, for our particular MF coupling, there is no net effect of NN interactions on the phase at the "sweet spot".

Third, the effects of geometry and of parameter strengths are independent. Since $t_{\uparrow}$ and $t_{\downarrow}$ are nonzero and unequal in most cases, we just need to focus on the equality $4f_{+}f_{-}+|h|^{2}=0$. As shown in Appendix B, this is equivalent to:

\begin{eqnarray}
|\cos(\vec{p}\cdot \vec{r}_{j}/2)|+|\cos(\vec{p}\cdot \vec{r}_{k}/2)| \ge |\cos(\vec{p}\cdot \vec{r}_{l}/2)|,
\label{eq11}
\end{eqnarray}
for any $(j, k, l)$ being a permutation of $(1, 2, 3)$.\\

\begin{figure}[tbp]
\centering
\includegraphics[width=0.35\textwidth]{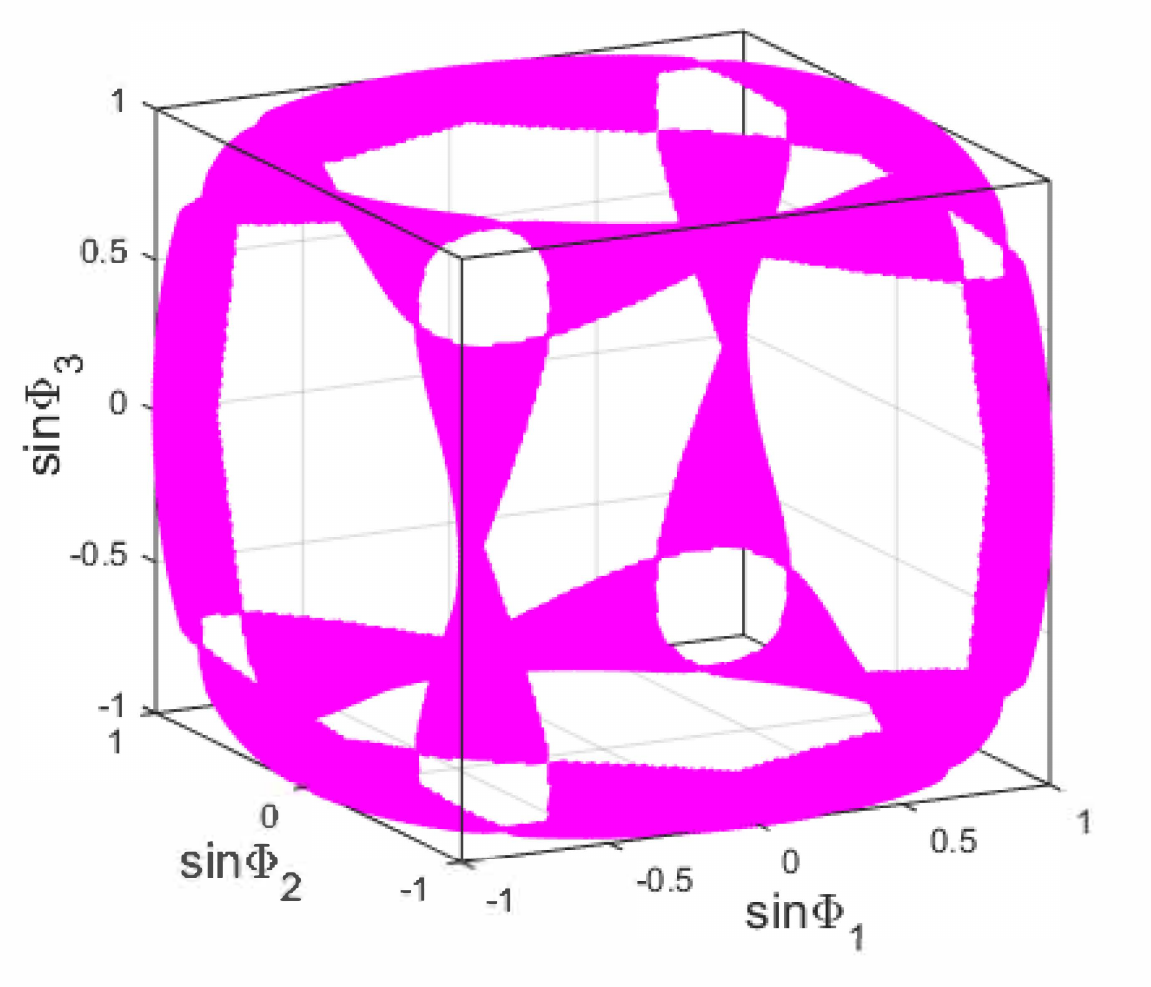}
\captionsetup{justification=raggedright}
\caption{Regions of the gapped SC phase in the 3-D parameter space $(\sin\Phi_{1}, \sin\Phi_{2}, \sin\Phi_{3})$. The shaded surface near the 12 edges of the cube shows the domain in which at least one inequality $|\cos\Phi_{j}|+|\cos\Phi_{k}| \ge |\cos\Phi_{l}|$ is violated, where $(j, k, l)$ is a permutation of $(1, 2, 3)$. This surface defines the gapped SC phase. Note that the only points in this parameter space that have physical significance are those for which $\Phi_{2}=\Phi_{1}+\Phi_{3}$ (including both the near-edge and kernel regions).}
\label{fig5}
\end{figure}

The above inequalities describe the parameter range of the gapless phase as compared to the gapped SC phase and actually give a measure of the "strength" of TRS breaking. The phase diagram in Fig. \ref{fig4a} obtained by numerical simulations of band structures at every point in $p_{x}-p_{y}$ plane is exactly the same with that obtained according to the three inequalities. Defining the Peierls phases associated with the complex NNN hoppings inside a hexagon as $\vec{p}\cdot \vec{r}_{j}/2=\Phi_{j}$ (with restriction $\Phi_{2}=\Phi_{1}+\Phi_{3}$), the distribution of gapped SC phase in the $(\sin\Phi_{1}, \sin\Phi_{2}, \sin\Phi_{3})$ parameter space is shown in Fig. \ref{fig5}. It can be seen that the gapped SC phase is mainly localized near the edge of the cube, which indicates that at least two of the three $|\sin\Phi_{j}|$ are near $1$. We know that $2t_{\sigma}\sin\Phi_{j}$ is the amplitude difference of NNN hopping terms before and after the TRS transformation. So we claim that the gapped SC phase corresponds to "strong" TRS breaking in which at least one of the three inequalities (\ref{eq11}) is violated. For "strong" TRS breaking, two of the three Peierls phases lead to a relatively large energy difference after the TRS transformation, and the energy gap is large enough to protect the MZMs from coupling with bulk modes. By contrast, the gapless phase corresponds to "weak" TRS breaking, which is supported by its occupying the central part near $(p_{x}=0, p_{y}=0)$ in the phase diagram. Thus, this measure further divides the TRS-broken class into two groups.

Moreover, the phase diagram showing "$w=0$ vs. $w=\pm 1$" (Fig. \ref{fig4b}) has 3-fold rotational symmetry. This phase diagram is obtained by numerically calculating~\cite{Chernnumdis} the winding number, $w$, for every point in $p_{x}-p_{y}$ plane:

\begin{eqnarray}
w=\frac{i}{2\pi}\int_{T^{2}}(\langle\partial_{k_{x}}u_{1}|\partial_{k_{y}}u_{1}\rangle-\langle\partial_{k_{y}}u_{1}|\partial_{k_{x}}u_{1}\rangle)d^{2}k,
\label{eq12}
\end{eqnarray}
where $|u_{1}\rangle$ is the eigenstate of band 1 and $T^{2}$ is the first Brillouin zone. Since the sum of the topological invariants of the neighboring band 1 and band 2 are zero, it's enough to do such a calculation for one band. In the pattern of the phase diagram, the center is mainly the topologically trivial region, $w=0$, while the surrounding $w=1$ (blue) and $w=-1$ (red) nontrivial regions are separated by "thin-ribbon" regions with $w=0$. The central lines of the six "thin-ribbon" green regions indicate $\vec{p}\perp \vec{r}_{j}$, for which one of the three Peierls phases is zero. The topological transition only occurs when the gap between band 1 and band 2 closes so that the topology of bulk bands intrinsically changes. The bulk bands only touch at the typical Dirac points $(K_{x}, \pm K_{y})$ ($K_{x}=\frac{2\pi}{3}$ and $K_{y}=\frac{2\pi}{3\sqrt{3}}$), which indicates the decoupling of the two sublattices. Band touching at only one of the two inequivalent Dirac points entails the change of winding number by $1$, while the case of touching at both hasn't been observed in this model. Additionally, when $w$ is nonzero, the system is characterized by ordinary gapless edge modes connecting band 1 and band 2 at two armchair edges. The interaction between the gapless edge modes and the Majorana flat bands will be the subject of the future study.

In general, when $\mu\neq 0$, the system displays changes in both of the gapped SC and gapless phases. When $|\mu|$ increases from zero in the gapped SC phase, the gap between band 1 and the zero-energy line gradually decreases until 2 MZMs couple with the bulk modes, leading to a decreasing jumping of the degeneracy of MZMs. The pair of fully pseudospin-polarized MZMs that is exactly localized at the first layer in $\mu=0$ case, will extend to the deeper layers with exponentially decreasing amplitudes. In the gapless phase, the pairing occurs in a partially filled band. The Dirac points connecting band 1 and band 2 vanish, accompanied by the relative displacement of the upper and lower Dirac cones. The gapless property is preserved but the bands are indirectly closed.

At last, it should be mentioned that our model has several remarkable differences from previous models for realizing MFs in 2-D optical lattices~\cite{MFquasi1D,MFEquil,1DTOPChain}. First, the geometric structure in our model is not one or more topological chains with transverse tunneling in a topological trivial background. It naturally includes the tunnelings in different directions (which is longitudinal/transverse in a square lattice, and is slanted in a honeycomb lattice).

Second, the commonly used ten-fold classification of fermionic phases~\cite{schnyder2008classification} may be not directly applicable to our model due to its dependence on $\vec{p}$ as a parameter. More mathematical tools may be needed in order to apply the classification rule, such as constructing new charge, parity and time-reversal symmetry operators, which will be addressed in our future research. In addition, the gapless edge modes in our model are reminiscent of quantum anomalous edge states~\cite{wang2013anomalous}. By contrast, most other models can be covered by the ten-fold classification description. They inherit the topological properties of the 1-D TRS broken chain and have invariants of type-$Z_{2}$, while the number of effective MFs diversifies with the number of chains being even or odd.

Third, our theoretical model seems to be ahead of the development of experimental techniques in fermionic optical lattices. Some of the necessary techniques have been well developed as described in the introduction. The biggest challenge may be the precise tuning of both the spin-singlet and spin-triplet pairings. Recent progress in implementing pairings in cold atoms includes finite-momentum Cooper pairings. These can induce Fulde-Ferrell (FF) states which have superconducting order parameters with uniform amplitudes but spatially dependent phases. Protocols for creating such FF superfluids in spinful~\cite{qu2013topological,liu2016detecting,dong2013fulde,xu2014competing,he2018realizing} and spinless~\cite{zheng2018fulde} cold atoms may be helpful for implementing the crucial Hamiltonian components in our model, though it is still a big step. Further experimental progress is needed to realize our model, but our study of this model Hamiltonian may be insightful for researchers in cold atom physics.

\section{Conclusion}

We have proposed a method for creating Majorana fermions at the edge of a honeycomb optical lattice of ultracold atoms. This is done by generalizing a 2-D topologically nontrivial Haldane model and introducing textured pairings. Both the spin-singlet and textured spin-triplet pairings are added to a pseudospin-state dependent lattice, whose time-reversal symmetry is broken by complex next-nearest-neighbor hoppings. If generalized "sweet spot" conditions are satisfied, Majorana zero modes will arise on a single edge of the lattice. We have analyzed their properties, such as pseudospin polarization. We find that this system has a gapped superconducting phase and a gapless phase, each of which can have zero or nonzero winding numbers, and have calculated the phase diagrams of the system. We have simplified the understanding of the bandgap-closing condition of the bulk Hamiltonian by identifying a discriminant that distinguishes the gapped superconducting and gapless phases, and provides a measure of the "strength" of time-reversal symmetry breaking. Further developments of this model may include interactions between Majorana fermions and the interaction between Majorana zero modes and ordinary gapless edge modes.

\section{Acknowledgments}

We thank Dong-ling Deng and Haiping Hu for helpful discussions. This material is based upon work supported by the US National Science Foundation Physics Frontier Center at JQI.

\appendix
\section{Hamiltonian in Majorana representation}

Here we show our Hamiltonian $\hat{H}$ expressed as a sum of products of Majorana operators, $\gamma$. The generalized "sweet spot" conditions are apparent in this representation. In our model, we choose to make the terms of $\gamma_{\vec{r}_{j},\uparrow,1}\gamma_{\vec{r}_{j^{\prime}},\uparrow,1}$, $\gamma_{\vec{r}_{j},\uparrow,1}\gamma_{\vec{r}_{j^{\prime}},\uparrow,2}$,
$\gamma_{\vec{r}_{j},\downarrow,2}\gamma_{\vec{r}_{j^{\prime}},\downarrow,2}$ and $\gamma_{\vec{r}_{j},\downarrow,2}\gamma_{\vec{r}_{j^{\prime}},\downarrow,1}$ vanish, which will lead to Eq.(\ref{eq4})-(\ref{eq6}).

\begin{eqnarray}
  \hat{H}=\frac{i}{2}\sum_{\langle j,m\rangle }\Big \lbrack (-t-\Delta)\gamma_{\vec{r}_{j},\uparrow,1}\gamma_{\vec{r}_{m},\downarrow,2}
  +(t-\Delta)\gamma_{\vec{r}_{j},\uparrow,2}\gamma_{\vec{r}_{m},\downarrow,1}  \Big \rbrack \nonumber \\
  +\frac{i}{2}\sum_{\langle \langle j,j^{\prime}\rangle \rangle, y_{j}< y_{j^{\prime}}}\sum_{\sigma=\uparrow,\downarrow}\Big \lbrack (-t_{\sigma}\sin\phi_{\sigma}+\textrm{Im}(\Delta_{\sigma}))\gamma_{\vec{r}_{j},\sigma,1}\gamma_{\vec{r}_{j^{\prime}},\sigma,1} \nonumber \\
  +(-t_{\sigma}\sin\phi_{\sigma}-\textrm{Im}(\Delta_{\sigma}))\gamma_{\vec{r}_{j},\sigma,2}\gamma_{\vec{r}_{j^{\prime}},\sigma,2} \nonumber \\
  +(-t_{\sigma}\cos\phi_{\sigma}-\textrm{Re}(\Delta_{\sigma}))\gamma_{\vec{r}_{j},\sigma,1}\gamma_{\vec{r}_{j^{\prime}},\sigma,2} \nonumber \\
  +(t_{\sigma}\cos\phi_{\sigma}-\textrm{Re}(\Delta_{\sigma}))\gamma_{\vec{r}_{j},\sigma,2}\gamma_{\vec{r}_{j^{\prime}},\sigma,1}
  \Big \rbrack  . \nonumber \\
\label{eq13}
\end{eqnarray}
\\

\section{Derivation of the phase boundary between gapped SC and gapless phases}

This section derives the expression for the phase boundary between gapped SC and gapless phases in the $p_{x}-p_{y}$ plane when other parameters in our model are fixed. The results are in close agreement with the phase diagram (Fig. \ref{fig4a}) that was obtained by numerical simulations of band structures for every point in $p_{x}-p_{y}$ plane. This supports the correctness of our mathematical analysis, exemplifies the bulk-edge correspondence in topological physics and provides the basis for our identification of "strong vs. weak" of TRS breaking.

According to the results of section $\textrm {\uppercase\expandafter{\romannumeral 4}}$, the band touching condition reduces to the vanishing of the discriminant given in Eq. (\ref{eq10}). We now show the steps of deriving the inequality (\ref{eq11}).

Let $z_{j}=e^{i\vec{k}\cdot \vec{r}_{j}}$, $a_{j}=\cos(\vec{p}\cdot \vec{r}_{j}/2)$, for $j=1,2,3$. Note that $z_{2}=z_{1}z_{3}$ since $\vec{r}_{2}=\vec{r}_{1}+\vec{r}_{3}$. For any $(z_{1}, z_{3})$ such that $|z_{1}|=|z_{3}|=1$, the corresponding vector $\vec{k}=(k_{x}, k_{y})$ can be determined. Thus, we have converted the problem of finding such a $\vec{k}$ to a problem of finding $z_{j}$. Using "$\Leftrightarrow$" to denote equivalence, it is easy to show that:\\

\begin{eqnarray}
\sum_{j=1}^{3}z_{j}a_{j}=0 \Leftrightarrow z_{1}=\frac{-a_{3}}{a_{1}z_{3}^{-1}+a_{2}}. \nonumber
\end{eqnarray}

\begin{eqnarray}
&&\textrm{So } |z_{j}|=1 (j=1,2,3) \nonumber \\
\Leftrightarrow &&\textrm{there exists a } z_{3} \textrm{ with } |z_{3}|=1  \nonumber \\
&&\textrm{ such that } |a_{3}|=|a_{1}z_{3}^{-1}+a_{2}|. \nonumber
\end{eqnarray}

Considering that $||a_{1}|-|a_{2}|| \le |a_{1}z_{3}^{-1}+a_{2}| \le |a_{1}|+|a_{2}|$ for $|z_{3}|=1$, the above requirement\\
\begin{eqnarray}
&&\Leftrightarrow ||a_{1}|-|a_{2}|| \le |a_{3}| \le |a_{1}|+|a_{2}| \nonumber \\
&&\Leftrightarrow |a_{1}|-|a_{2}| \le |a_{3}|, |a_{2}|-|a_{1}| \le |a_{3}|, |a_{3}| \le |a_{1}|+|a_{2}|.  \nonumber
\end{eqnarray}\\

Therefore, the band touching condition is finally reduced to:

\begin{eqnarray}
|a_{j}|+|a_{k}| \ge |a_{l}|,  \nonumber
\end{eqnarray}
for any $(j, k, l)$ being a permutation of $(1, 2, 3)$. This is the inequality (\ref{eq11}). Any special case with singularities can be verified to satisfy these conditions.

\bibliographystyle{apsrev4-1}
\bibliography{majoranafermion}

\end{document}